\documentclass{aa}

\usepackage{graphicx}
\usepackage{subcaption}         
\usepackage{lscape}             
\usepackage{placeins}           
                                
\usepackage{natbib}
\usepackage[hidelinks]{hyperref}
\hypersetup{
     colorlinks   = true,
     linkcolor    = blue,
     citecolor    = blue
}

\usepackage[T1]{fontenc}

\usepackage[varg]{txfonts}
\usepackage{graphicx}	
\usepackage{amsmath}	
\usepackage{amssymb}	
\usepackage{newtxtext,newtxmath}
\usepackage[normalem]{ulem}

\newcommand{\be}{\begin{equation}}
\newcommand{\ee}{\end{equation}}
\newcommand{\beal}{\begin{aligned}}
\newcommand{\eeal}{\end{aligned}}

\usepackage{esdiff}
\usepackage{array}
\usepackage{booktabs}
\usepackage{xcolor}
\usepackage{soul}
\usepackage{bbold}
\usepackage{multicol}

\begin{document}
\nolinenumbers
\title{Sculpting the outer edge of accretion disks in  pre-circumbinary\\ binary black hole systems}

\author{
Fabien Casse\inst{1}\thanks{E-mail: fcasse@apc.in2p3.fr}
\and Peggy Varniere\inst{1,2} 
\and L\'ena Arthur\inst{1} 
\and Fabrice Dodu\inst{1} }
\institute{Universit\'e Paris Cit\'e, CNRS, Astroparticule et Cosmologie, F-75013 Paris, France
\and Universit\'e Paris-Saclay, Universit\'e Paris Cit\'e, CEA, CNRS, AIM, 91191, Gif-sur-Yvette, France}

\date{Accepted XXX. Received YYY; in original form ZZZ}

\abstract
 {Binary black hole systems (BBHs) have become a vivid reality in astrophysics as {stellar-mass} black hole mergers are now detected through their related gravitational wave emission during the merger 
 stage. If many studies were recently dedicated to the last stages of BBH where black holes are surrounded by a circumbinary disk (CBD), the structure of these systems prior to the formation of the 
 CBD remains mostly unexplored.}
   {The aim of the present article is to investigate the potential modifications induced by the presence of a secondary black hole onto the structure of the accretion disk surrounding the primary black hole. Identifying potential specific features of the accretion flow in pre-circumbinary BBH may help us to identify such systems through their electromagnetic emission.} 
   {We performed 2D classical hydrodynamical simulations of an accretion disk surrounding the primary black hole while taking into account all gravitational effects induced by  both the primary black hole and the secondary black hole orbiting on circular orbits around the center of mass of the system.}
   {We report three main effects of the presence of a secondary black hole orbiting a circular orbit beyond the outer edge of the accretion disk: 1/ the outer radius of the accretion disk is significantly reduced and its ratio to the black hole separation is directly linked to only
   the mass ratio of the black holes; 2/ two spiral arms are visible in the gas density structure of the disk and 3/ the outer edge of the accretion disk exhibits an elliptical shape that mainly depends on the mass ratio of the black holes.}
   {Our results show that an accretion disk orbiting a primary black hole in a pre-CBD BBH exhibits specific features induced by the gravitational force generated by the presence of a secondary black hole beyond its outer edge. Such features, directly linked to the binary separation and mass ratio, has therefore the potential to help in the search and identification of
   BBH in the pre-CBD stage. }


\maketitle

\section{Introduction} 

For more than three decades, binary black holes have remained a  conjecture in astrophysics when considering colliding galaxies hosting supermassive black holes \citep{Begelman80}. In that context, supermassive black holes are expected to bind gravitationally and to progressively get closer until they merge. Over time, numerous studies have been performed in order to understand what are the physical processes that can preside such in-spiral motion. For orbital separation beyond the parsec scale the dynamical friction of black holes with the galactic gaseous environment seems to be the leading process \citep{Mayer07}. At orbital separation below $10^{-2}$ pc, gravitational wave emission is expected to be the key phenomenon driving the in-spiral motion until the fusion of the two compact objects. Between these two separation scales no clear consensus has emerged identifying the mechanism driving the in-spiral motion - the \lq final parsec problem\rq  - as three-body stellar interaction (see e.g. \citealt{Preto11}) or disk gas interaction (e.g. \citealt{Armita02,Cuadra09}) are candidates for this role. \\
Since the first detection of gravitational waves emitted by the fusion of  a stellar binary black hole \citep{Abbott16}, the binary black hole conjecture has become a vivid reality in the astrophysical community. Such discovery has enhanced the relevance of numerical studies devoted to the structure 
of the binary system, including the interaction between the two compact objects and the circumbinary disk (CBD) surrounding them \citep{macfayden08,Ragusa16,Munoz20,Siwek23,Franchini23}, the impact of the binary gravitational field upon the CBD \citep{Noble12,Zilhao15,Noble21,Mignon23}, the presence of individual accretion structures around each black hole (also dubbed as 'mini-disks')  \citep{Pihajoki18,Ingram21,Davelaar22} or the existence of the so-called 'lump' structure at the inner edge of the CBD \citep{Shi12,Noble12,dOrazio13,Mignon23b}. \\
Understanding the dynamical structure of the gas orbiting within the complex gravitational field generated by the binary black hole is likely to be a key to discriminate between single black holes and pre-merger binary black holes through the electromagnetic output of the system. 
So far the characterisation of electromagnetic signatures of binary black hole systems  have been performed in the framework of black hole orbital separation up to a few hundreds of gravitational radii (see e.g. \citealt{Tanaka12,Tang18,Dascoli18,Combi22,Major23,Cocchia24}) and suggest that electromagnetic binary black hole signature should be detectable  
{for most of the mass ratio of the binary black-holes} provided the redshift of these systems remains relatively small.  
It is noteworthy that all the aforementioned studies have been performed assuming ad-hoc initial conditions that are not originating from simulations
 devoted to earlier phases of the binary (see e.g. \citealt{Lai23} for a review on general binaries and \citealt{Duffell24} for a review on numerical simulations devoted to supermassive BBH). In 
 particular it is still unclear how the CBD forms around the binary black holes thus leading to uncertainties regarding initial conditions used in fluid simulations and the {derived} electromagnetic 
 emission of the binary in the CBD stage. Simulations of the early phase of the binary, namely before the formation of the CBD, are then required in order to bridge the gap between the very early 
 stage ('pre-CBD') and the CBD stage of the binary system. The transition between these two stages of BBH systems is not yet clearly identified in terms of black hole separation.
  However since no active galactic nuclei (AGN) do exhibit accretion disks whose outer radii are larger a few thousands of gravitational radii of the central black hole
 \citep[see e.g.][and references therein]{Jha21} 
  it is likely that CBDs do not stretch to more than a few thousands gravitational radii. As such, the separation between the two black holes in the CBD stage is likely to be of the order of, at most, a few hundreds of gravitational radii. Regarding pre-CBD BBH, the separation between the two black holes can be quite large and even larger than the typical size of AGN disks so one can assume that, in the pre-CBD stage, one or even two of the black holes are still surrounded by their own 'primordial' accretion disks.   
  
\noindent Pre-CBD BBH are exhibiting gravitational fields that are quite similar to other binary systems such as binary stars or star-planet systems. In a 
 classical gravitational context, numerous authors have already investigated the impact of the presence of a bound gravitational disruptor upon the accretion disk surrounding a central star or 
 planet. Among the various particularities of these disks, the various studies agree on that the radial extent of the disk is limited well below the Roche lobe size. Such result was obtained using 
 various approaches such as  exploring test particles periodic orbits  \citep{Paczynski77,Eggleton83,Pichardo05}, applying perturbation theory to hydrodynamical disk equilibrium 
 \citep{Papaloizou77,Artymowicz94,MirandaLai15} or using smoothed-particle hydrodynamics simulations \citep{Artymowicz94,MartinLubow11}.  Regarding the internal structure of individual 
 disk, the presence of the secondary object is likely to induce spiral waves propagating in the disk surrounding the primary black hole as shown by \citet{Sawada86} and \citet{Spruit87}. Other 
 studies have been performed in various astrophysical contexts and support the existence of such waves for instance in Cataclysmic Variables \citep{Godon98,Ju16}, protoplanetary disks 
 \citep{Rafikov16,Bae18}, circumplanetary disks \citep{Zhu16} and in mini-disks formed around components of BBH systems \citep{Ryan17}. Finally, the gravitational influence of the secondary 
 component of the binary system is also believed to affect the outer edge of the disk orbiting the primary component leading to an elliptical shape \citep{Whitehurst91,Lubow94,HD100546,Wester24}.  
 Based on the aforementioned studies and the similarities regarding gravitational fields, one would expect similar features to also appear in individual accretion disks orbiting black holes in pre-CBD BBH systems. 
 
\noindent The goal of the present paper is to make a first step upon the bridge between pre-CBD systems and BBH in CBD stage by studying the gravitational impact of a secondary black hole upon an accretion disk orbiting around a primary black hole while assuming both black holes to be gravitationally bound. The existence of such link would help discriminate among BBH candidates and could potentially lead to the identification 
of pre-CBD BBH systems. The aim of such simulations is to assess if it is possible to link the characteristics of the disk surrounding the primary black hole, hence its observables, with the parameters of the binary, namely its separation and mass ratio.  Such detection would be invaluable as it would provide observational clues to the gas structure throughout the evolution of the binary and thus to the formation of the CBD.\\

\noindent The paper is organized as follows: the second section is devoted to the numerical framework and setup of simulations of a large accretion disk surrounding a primary black hole  (hereafter dubbed as the  circumprimary disk) and prone to the gravitational influence of a secondary black hole orbiting beyond the outer edge of the accretion disk. The third section presents the results of equal-mass BBH  hydrodynamical simulations while the fourth section generalizes the results to unequal mass BBH systems. The last section summarizes the results and discusses the perspectives of our work.    

\section{Hydrodynamic simulations of accretion disks in early BBH systems}

In this paper we aim at studying the effects of the presence of an external gravitational disruptor upon the accretion disk orbiting around a black hole. As we focus our study on the early phase of the BBH, namely when the distance between the black hole and the secondary object $D_{12}$ is much larger than the size of the black hole gravitational radius $r_{\rm g1}=GM_1/c^2$. In the previous expression $G$ stands for the gravitational constant while $c$ is the velocity of light. The other important parameter describing the BBH is obviously the mass ratio of the two black holes that we define as $q=M_2/M_1$  {with $M_1$ the heavier black-hole}.

\subsection{Early stage of circular orbiting BBH: the pre-CBD stage}
The two black holes composing the BBH are gravitationally bound and orbit around the center of mass of the system. Many studies have focused on the impact of mass and angular momentum transfer between the CBD and the binary and found that the trajectories of the two objects display eccentricity whose values is ranging between  $0.3$ and $0.8$, except if they are initially close to a circular orbit 
\citep{Armi05,Cuadra09,Roedig11,Munoz19,Zrake21}. Such result is valid in the context of a BBH surrounded by a CBD and does not include the momentum loss driven by the emission of gravitational waves.
 For earlier systems, namely pre-CBD BBH, no firm constraint has yet been put on the motion of the two black holes so we chose the simplest configuration possible, namely having both black holes 
 moving on circular orbits around the center of mass of the BBH in the same plane as the circumprimary accretion disk.\\   
 If one assumes that the loss of angular momentum in the binary is mainly induced by the emission of gravitational waves, the variation of the orbital separation of the binary $D_{12}$ occurring over a timescale $dt$  is provided by \citep{hughes09}
\begin{equation}
\frac{dD_{12}}{D_{12}}=-\frac{64G^3(M_1+M_2)^2\mu}{5c^5D_{12}^4}dt
\label{Eq:shrink}
\end{equation}
where $\mu=M_1M_2/(M_1+M_2)$. 
In a circular orbit configuration, the angular frequency of the BBH is given by
\begin{equation}
\Omega_{\rm BBH} = \left(\frac{G(M_1+M_2)}{D_{12}^3}\right)^{1/2}
\end{equation}
while the distances of each object from the center of mass of the system are $r_1 = q\  D_{12}/(q+1)$ and $r_2 = D_{12}/(q+1)$ respectively. The velocity of both the black hole and the secondary object in the frame of the center of mass of the system will then be
$\beta_1^2=V^2_1/c^2=q\ r_{\rm g1}/D_{12}$ and $\beta^2_2 = r_{\rm g1}/D_{12}$. Early stages of the BBH will then lead to black hole velocities much smaller than the velocity of light, thus opening the way to a classical description of the motion of both black holes. In
the rest of this paper we will use the orbital period of the BBH as time reference, namely
\begin{equation}
P_{\rm BBH} = 2\pi\left(\frac{D^3_{12}}{GM_1(1+q)}\right)^{1/2}=\frac{2\pi r_{\rm g1}}{c}\left(\frac{D^3_{12}}{r^3_{\rm g1}(1+q)}\right)^{1/2} \ .
\label{eq:PBBH}
\end{equation}

\noindent In order to fulfill the circular orbit assumption made in this paper, one has to check that the 
shrinking of the binary separation
 is limited over one binary period, for instance $|\Delta D_{12}/D_{12}|\leq 0.1\%$ for $\Delta t=P_{\rm BBH}$. The previous equations then lead to a lower limit of the binary separation $D_{12}$, namely
\begin{equation}
\left|\frac{\Delta D_{12}}{D_{12}}\right| = \frac{128\pi q(1+q)^{1/2}}{5}\left(\frac{r_{\rm g1}}{D_{12}}\right)^{5/2}\leq 0.1\%
\label{Eq:eqarlyBBH}
\end{equation}
which is verified if $D_{12}\geq 100\ r_{\rm g1}$. In order to stay well outside of the CBD forming phase we decided to restrict ourselves to separations $D_{12}\geq 500\ r_{\rm g1}$ which ensures that the BBH system is in the pre-CBD phase and that we can safely assume that the in-spiral motion due to gravitational wave emission is negligible over a very large amount of BBH periods. 
This is of prime interest as the follow up of the evolution of the accretion disk prone to an external orbiting massive object requires performing simulations of 
the fluid into the complex gravitational field generated by the presence of the two massive objects over a very long time, namely up to a thousand binary periods. \\ 
While the secondary black-hole is probably surrounded by its own accretion disk, the gravitational influence this disk would have upon the 
circumprimary disk is likely negligible compared to the effect we are studying here so we did neglect this force in our simulations. We also did not include the secondary black hole and its accretion disk in our 
computational domain as we chose to focus on the circumprimary disk which is likely to be the most visible gas structure in the BBH as we set the primary black hole to be the most massive object of the binary ($q\leq 1$).
In our study we will then consider two parameters of the system as free, namely the binary separation $D_{12}$ and the black hole mass ratio $q\leq 1$. Altogether these two free parameters characterized the 
gravitational influence of the secondary black hole upon the circumprimary accretion disk provided the orbits of the black holes are circular. 
The parameters for the various simulations performed in this study are listed in  
Tab.\ref{Tab_all}.

\subsection{Hydrodynamics simulations of the accretion disk around black hole in BBH}
\subsubsection{Simulation framework}
The best frame to perform our simulations in is the one coinciding with the primary black hole as it is possible to ensure in that frame the numerical conservation of the angular momentum of the gas 
(this frame will be called $\cal{R}$ hereafter). In such frame the center of mass of the accretion disk remains at a constant distance from the primary black hole but one has to consider fictitious forces 
at work upon the fluid. Indeed, the frame associated with the primary black hole has a circular motion around the center of mass of the BBH system. In the cartesian frame associated with the center 
of mass, we choose to keep the axis of $\cal{R}$ parallel to the axis of the inertial frame of the center of mass of the system. 
This ensures that the frame $\cal{R}$ is not rotating so that no Coriolis forces have to be accounted for in this frame. As a result the secondary black hole will also experience a circular 
motion in $\cal{R}$ whose radius and angular velocity are $D_{12}$ and $\Omega_{\rm BBH}$ respectively.   

\noindent Within our set of simulations we aim at describing the long-term evolution of the entire disk orbiting around the primary black hole. In order to do so we need to include the whole extent of the disk 
from the innermost stable circular orbit (ISCO) to the external radius of the disk. As the gravity of the primary black hole closely matches Newtonian gravity beyond $20\ r_{\rm g1}$ we choose to use a classical 
hydrodynamics framework. This choice enables us to keep good accuracy in our description of the disk while optimizing computing time. Moreover, encompassing the gravitational effects of the secondary 
black hole will be much easier as its influence is of Newtonian nature as any massive object would be {at such distance}.\\
BBH systems in their earliest phase are slow evolving systems so the approach of the two black holes and the mutual gravitational interaction is very likely to align accretion disk spin with the plane 
hosting the motion of the black holes. Accordingly we will consider 2D simulations within that plane (see e.g. \citealt{Liska21}). In order to take into account the relativistic nature of the primary black hole 
we use the pseudo-Newtonian  description provided by \cite{paczynsky_thick_1980} so that the inner edge of the disk will form near the ISCO. We chose this description as we aim at studying the global 
behavior of the disk with an emphasis on its outer parts where the spin of the primary black hole is likely to have a negligible impact.

\subsubsection{Hydrodynamics equations}

The set of hydrodynamics equations translates the conservation of mass and momentum of the gas in the complex gravitational field of the BBH system. These conservation laws write
\begin{eqnarray}
\frac{\partial \rho}{\partial t} + \nabla\cdot(\rho\mathbf{u}) &=& 0 \nonumber \\
\frac{\partial \rho\mathbf{u}}{\partial t} + \mathbf{\nabla}\left(\rho\mathbf{u}\mathbf{u} + P\mathbb{1}\right) &=& \mathbf{f}_{BH_1}+\mathbf{f}_{BH_2}+ \mathbf{f}_{\cal{R}}
\label{Eq:hydro}
\end{eqnarray}
where $\rho$ is the gas density, $\rho\mathbf{u}$ is the gas momentum density, $P$ is the thermal pressure and $\mathbb{1}$ is the identity matrix. The momentum conservation law takes into account the gravitational force densities generated by the central black hole using a pseudo-Newtonian prescription of the force of the central black hole 
\begin{equation}
\mathbf{f}_{BH_1} = -\rho c^2r_{\rm g1}\frac{\mathbf{r}}{r(r-2r_{\rm g1})^2}
\end{equation} 
where $\mathbf{r}$ is the position vector of a given gas element in the frame of the central black hole where $\lVert\mathbf{r}\rVert=r$.
 The gravitational force density originating from the secondary black hole $\mathbf{f}_{BH_2}$ orbiting around the primary one in the $\cal{R}$ frame can be described with very good accuracy by a Newtonian law as the distance between any element of gas and the secondary black hole  remains large compared to the gravitational radius $r_{\rm g2}$ of the second black hole. This force reads
 \begin{equation}
\mathbf{f}_{BH_2} = q\rho c^2r_{\rm g1}\frac{\mathbf{D}_{12}-\mathbf{r}}{\Vert\mathbf{D}_{12}-\mathbf{r}\rVert^3}     
 \end{equation}
 where $\mathbf{D}_{12}$ is the position vector of the second black hole in the frame of the central black hole, namely the $\cal{R}$ frame.
 In addition to these two gravitational force densities, one also has to consider the fictitious forces due to the rotation of $\cal{R}$. As already mentioned, the orientation of the axis of the $\cal{R}$ frame remains constant so no Coriolis force has to be considered and the only fictitious force is then 
 \begin{equation}
     \mathbf{f}_{\cal{R}} = - q\rho c^2r_{\rm g1}\frac{\mathbf{D}_{12}}{D_{12}^3}
 \end{equation}
 In order to close the set of equations we use a polytropic equation of state providing the value of the thermal pressure $P$ as a function of the density, namely $P=\kappa\rho^\gamma$ with $\gamma=5/3$ and $\kappa=10^{-4}$. The value of $\gamma$ is chosen in agreement with monoatomic gas characteristics. 
\subsubsection{Numerical simulation setup} 
The goal of this paper is to study the impact of the gravitational influence of a secondary black hole upon the circumprimary disk.
In contrast to single black hole systems, the presence of a second black hole in the BBH system likely 
impacts the outer edge to the accretion disk of the primary black hole.
In order to quantify that effect, we set up hydrodynamics simulations solving the aforementioned equations  in the 2D orbital plane. The simulation domain is designed to {encompass}  
the entire accretion disk from its inner to its outer edges. The simulation domain is thus set to be $\phi\in[0,2\pi]$ and $r\in [5r_{\rm g1},0.98D_{12}]$ so that this domain contains the inner edge of the accretion disk while maintaining velocities below the speed of light. The radius $R_{L_1}$ of the $L_1$ Lagrangian point of the BBH system is approximately \citep{plavec64,hameury01}
\begin{equation}
    \frac{R_{L_1}}{D_{12}} =\frac{1}{2} - 0.227\log_{10} q
\end{equation}
so that it lies within our simulation domain for any mass ratio $q\geq 7\times 10^{-3}$. The location of the $L_1$ Lagrangian 
point of the system being a helpful indicator of the upper limit of the outer radius of an accretion disk in a binary system, we 
thus make sure that the evolution of the outer edge of the accretion disk will be captured by our simulation.\\ 
The initial conditions of the simulations are consistent with the balance of an accretion disk prone to the gravitational field of a single non-spinning black hole, namely with an inner edge located near the innermost stable circular orbit (ISCO)  with $r_{\rm isco}=6r_{\rm g1}$ and an outer edge located close to the outer limit of the simulation domain. The initial gas density is prescribed as 
\begin{equation}
\rho(r) = \frac{\rho_o}{4}\left(1+\tanh{\frac{r-r_b}{\sigma_b}}\right)\left(1-\tanh{\frac{r-r_{\rm ext}}{\sigma_{\rm ext}}}\right)\left(\frac{6r_{\rm g1}}{r}\right)^{3/2}
\end{equation}
where $\rho_o=10$, $r_b=9 r_{\rm g1}$ and $r_{\rm ext}=0.7D_{12}$  \citep[see][]{Vin13}. The two constants $\sigma_b$ and $\sigma_{\rm ext}$ have been set to $0.9r_{\rm g1}$ and $0.05D_{12}$ respectively in order for the disk to exhibit  edges where density drops to a floor value $\rho_{\rm min}$ which is ten orders of magnitude smaller than the maximal initial density value. Radial velocity is set to zero initially and the azimuthal velocity is computed in order for the gas to reach a full balance between the central black hole gravitational field, the centrifugal force and the thermal pressure of the gas. We have chosen the value of the parameter $\kappa$ in the polytropic equation of state to be $\kappa=10^{-4}$ in order to mimic the physics of a thin accretion disk. Indeed, this parameter is directly connected to the disk sound speed and to the disk scale height $h$ through the vertical hydrostatic equilibrium of the disk. According to the initial disk density, the pressure in the disk is consistent with a constant disk aspect ratio $h/r \sim 0.07$. 

\noindent The challenge of these simulations stems from the fact that there is a huge contrast between the dynamical timescale  of the gas lying at $r_{\rm isco}$ and the timescale related to the orbit of the secondary black hole. Indeed the ratio of these timescales is proportional to the ratio of the velocity of the secondary black hole to the velocity of the gas at $r_{\rm isco}$, namely
\begin{equation}
 \frac{V_{\rm BH_2}}{V_{\rm isco}} \simeq \sqrt{\frac{6r_{\rm g1}}{D_{12}}}\ll 1
\end{equation}
\begin{figure}[t]
\centering
\includegraphics[width=8.5cm]{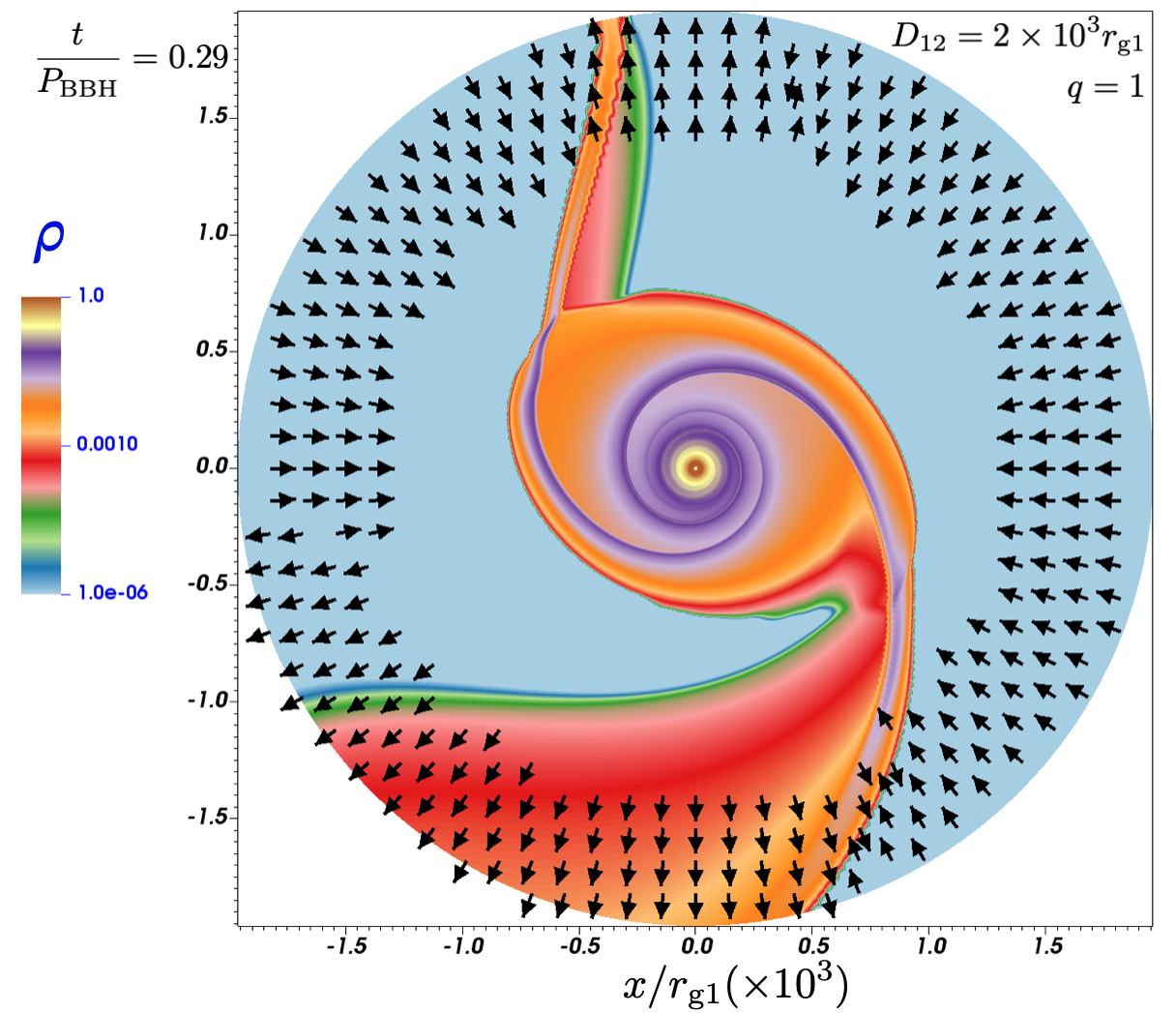}
\caption{Illustration of the outer boundary conditions imposed on the computational domain with the colormap representing the density while arrows show the sign of the radial velocity of the matter in the outer part of the computational domain at time $t=0.29P_{\rm BBH}$. On one hand gas is allowed to exit the computational domain without modifying its density while on the other hand the inflowing gas can enter the computational domain with a density set to a value very close to the density floor. Such conditions are designed to  mimic the ambient medium standing outside of the disk and to prevent any artificial input of gas from the exterior of the domain. }
\label{fig:Cond_init}
\end{figure}

\begin{figure*}[th]
\begin{tabular}{cc}
\includegraphics[width=8.5cm]{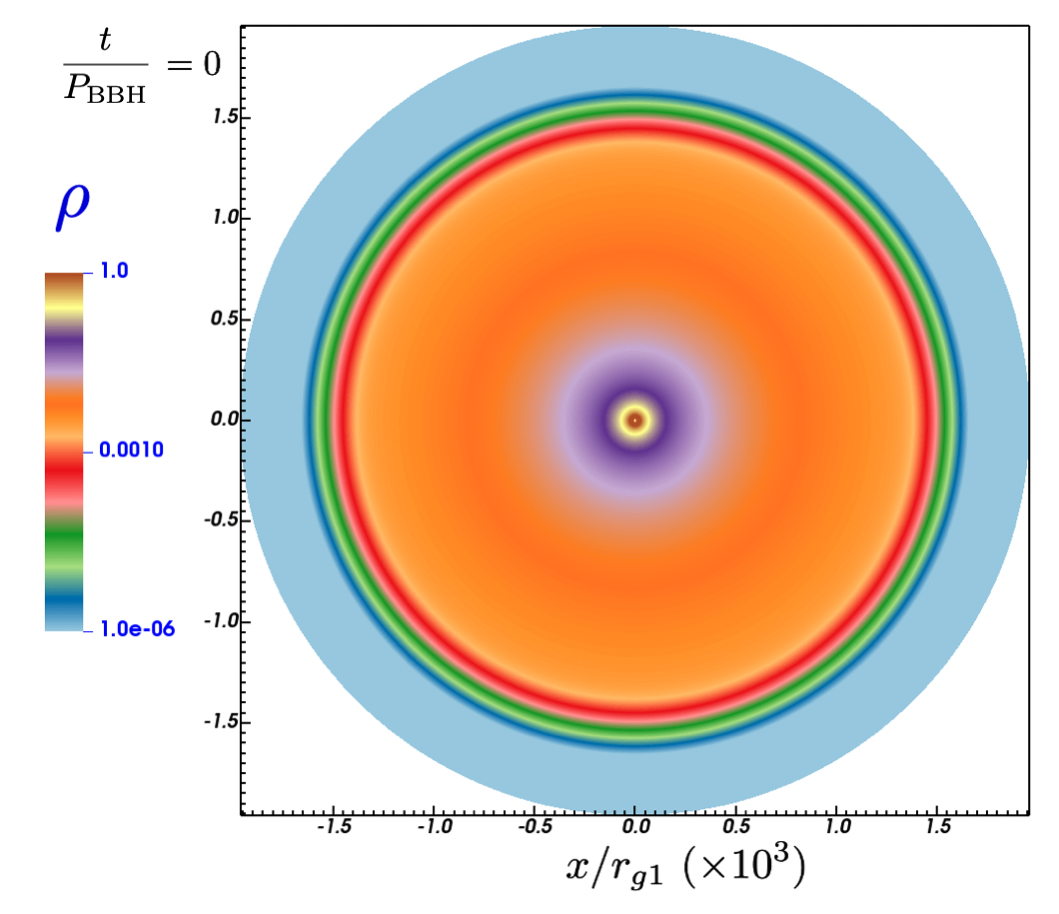} & 
\includegraphics[width=8.5cm]{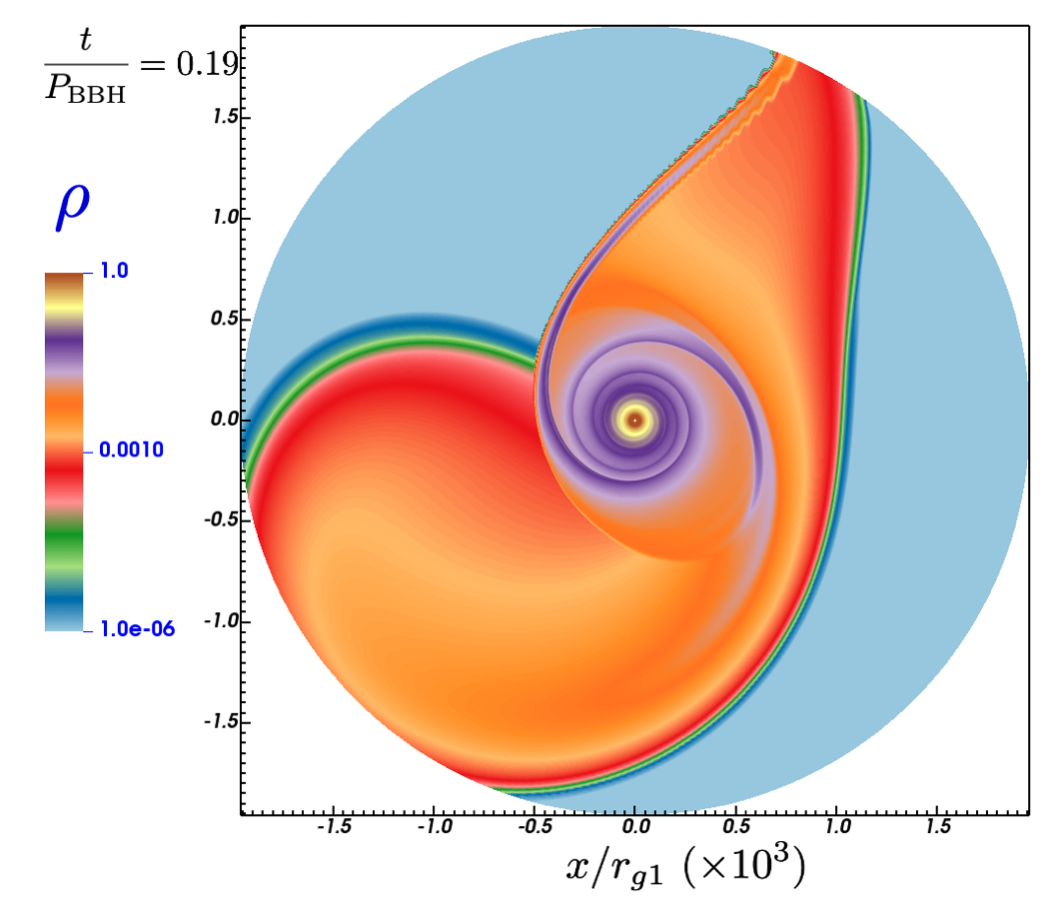} \\
\includegraphics[width=8.5cm]{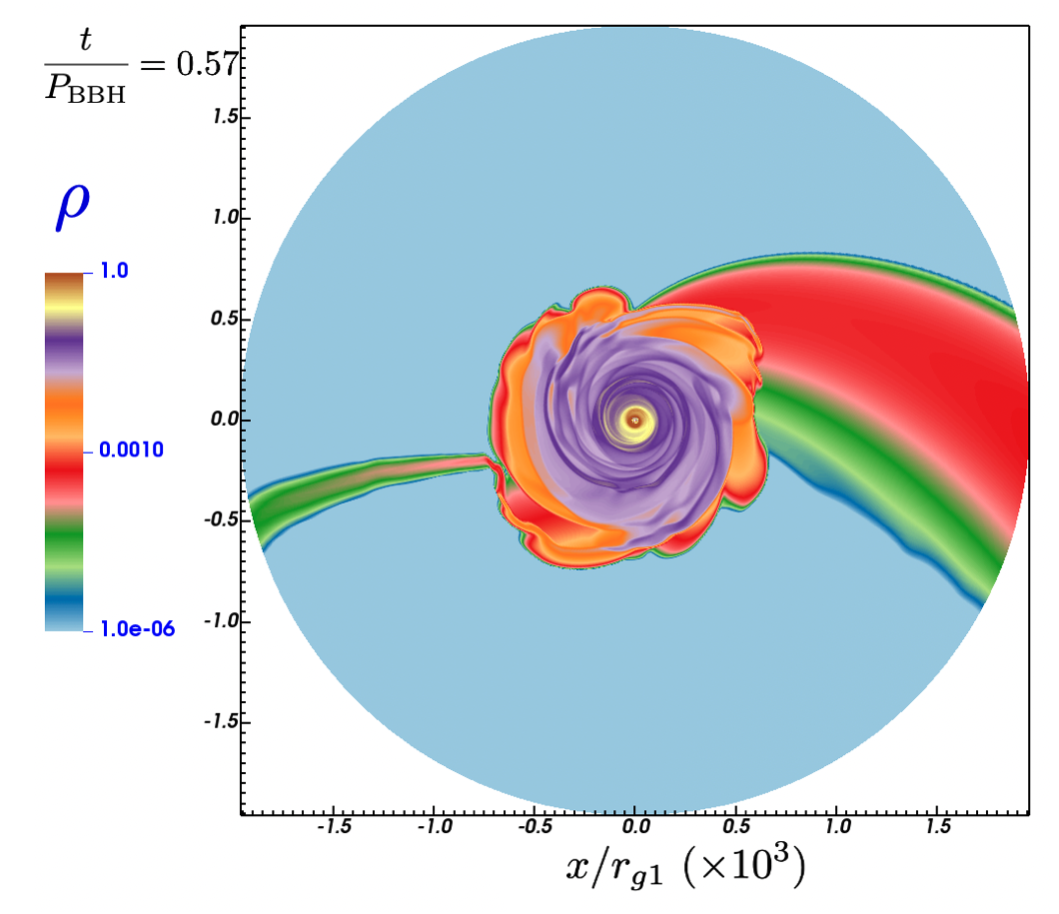} &
\includegraphics[width=8.5cm]{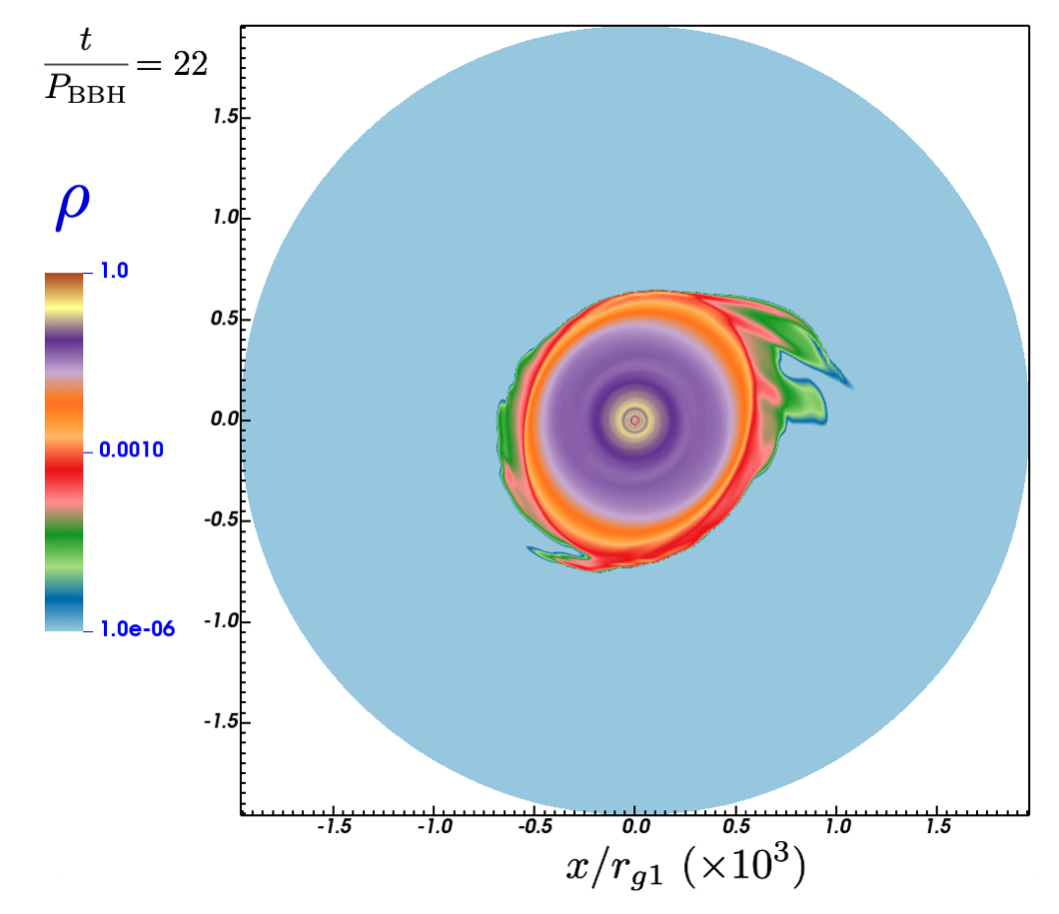}
\end{tabular}
\caption{Snapshots of the logarithmic density  of an accretion disk surrounding the primary black hole of a pre-CBD BBH. The secondary black hole orbiting in a circular motion around the primary black hole (in the frame of the primary black hole) has an equal mass  ($q=1$) while being separated by $D_{12}=2000\ r_{\rm g1}$.  Time is here normalized to the BBH period $P_{\rm BBH}$. The tidal effects induced by the presence of the secondary object leads to the removal of the outer part of the initial disk whose outer edge lies beyond the Lagrangian $L_1$ point of the BBH. After more than one orbital period of the BBH, the disk remains stable with an outer edge reduced to approximately $600\ r_{\rm g1}$ while displaying an elliptic outer edge shape whose eccentricity is  $e = 0.52 \pm 0.12$.}
\label{fig:Density_evol}
\end{figure*}
\noindent We should also stress that the spatial resolution needed to accurately describe the dynamics of the gas requires much smaller cells near the inner edge than at the outer edge of the simulation. In order to alleviate the numerical cost of these simulations we use a logarithmically radially spaced grid with $N_r=720$ cells in the radial direction and $N_\theta=320$ in the azimuthal one. The use of a logarithmic grid is an important asset to perform these simulations but cannot prevent us from considering a huge numerical cost for these simulations. Indeed, the time step of hydrodynamics simulations is controlled by the so-called CFL condition (Courant-Friedrish-Levy) translating the causality of hydrodynamical phenomena as 
\begin{equation}
\Delta t < \text{min}\left(\frac{\Delta X_i}{V_i}\right)
\end{equation}
where $\Delta X_i$ is the size of a cell in the $i^{th}$ direction.
The dominant velocity component of the gas in the accretion disk being the azimuthal one (with velocity is close to the Keplerian speed), the previous condition leads to a simulation time-step  
\begin{equation}
    \Delta t_{\rm isco}  \sim \frac{P_{\rm BBH}}{N_{\theta}}\left(\frac{6r_{\rm g1}}{D_{12}}\right)^{3/2}
\end{equation}
where $N_{\theta}=320$ is the azimuthal resolution of the grid. As we wish to follow the evolution of the whole accretion in pre-CBD binaries whose separations are up to $6\times 10^3\ r_{\rm g1}$, the previous relation indicates that we have to perform up to $10^7$ hydrodynamical iterations per BBH period $P_{\rm BBH}$.  
As an example, the number of iterations required to perform our  most remote binary black hole configuration is near ten billion time steps.\\
Regarding the boundaries of the computational domain, we opted for periodic boundary conditions in the azimuthal direction while outflowing conditions were imposed at the inner radial boundary. At the outer radial boundary of the simulation, we did not prevent gas from entering  the computational domain as this can lead to unphysical effects such as an artificial thermal pressure gradient. Instead we allow gas to inflow into the computational domain with a density very close to the floor density. This is especially valid as the outer edge of the disk stays well inside the grid while 
cells near the outer boundary are mimicking the ambient medium outside the disk. We have displayed on Fig.\ref{fig:Cond_init} an illustration of the outer boundary conditions: matter reaching the outer boundary can exit the domain (outer oriented arrows) while entering mass  (inner oriented arrows) is set to a density much lower than the one from the disk.\\
The numerical setup presented in this section has been used identically in all simulations displayed in this study. The entire setup is only dependent on the two BBH parameters, namely the
separation $D_{12}$ between the two black holes and the mass ratio $q$. 
As we express all physical quantities using the two aforementioned parameters we can then scan the range values of these parameters while ensuring an identical initial setup. 

\section{Impact of the binary separation on the circumprimary disk of an equal-mass BBH}

In order to study how the presence of a secondary black hole impacts the circumprimary disk, we choose to first focus on exploring how the separation between the two black holes impacts the circumprimary disk 
in the case of an equal mass pre-circumbinary BBH ($q=1$) configuration where we expect the  
most striking consequences.

\subsection{Sculpting the outer edge of a circumprimary disk in an equal-mass BBH: the fiducial case}

\label{equalBBH}
\begin{figure}
    \centering
    \includegraphics[width=0.48\textwidth]{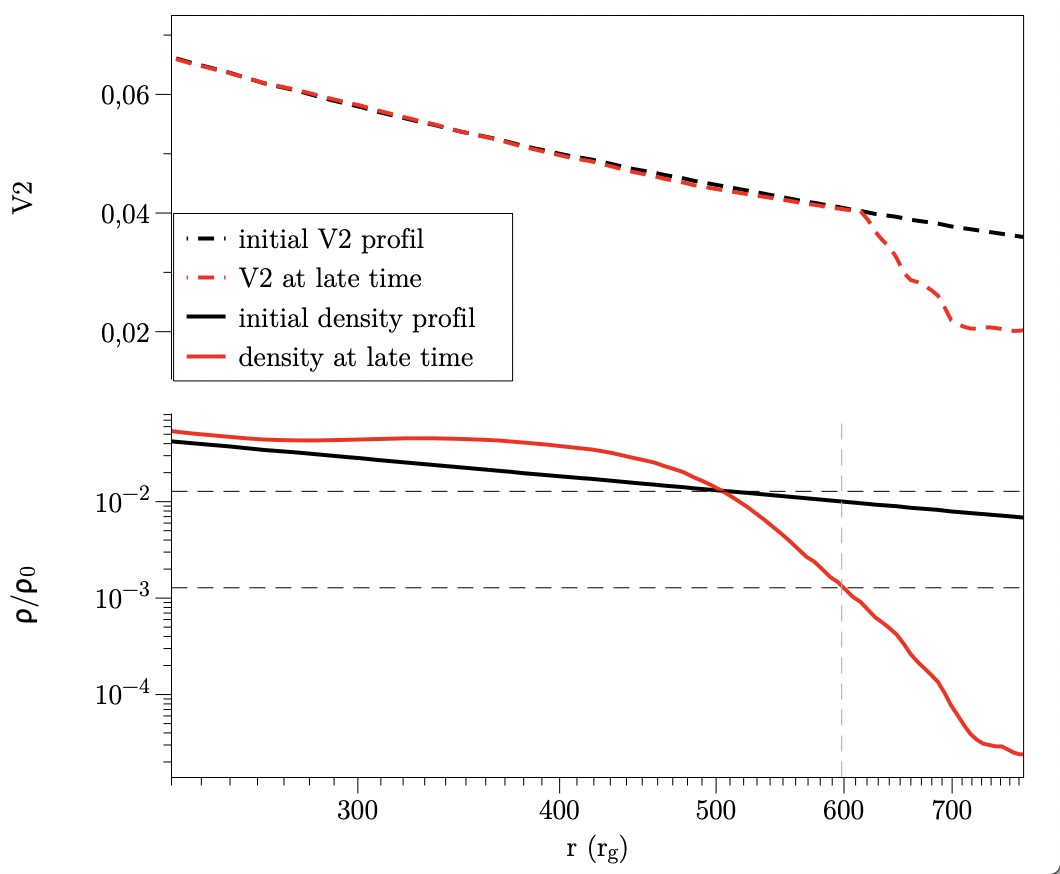}
    \caption{Comparison of the initial radial profiles of the density (upper panel) and  azimuthal velocity  (lower panel) of the disk with final profiles of the simulation. The profiles are obtained by azimuthally averaging the corresponding snapshots.}
    \label{fig:Profile_vphi_rho}
\end{figure}

We first present the fiducial case of a separation   $D_{12}=2\times 10^3r_{\rm g1}$, which is well above the threshold for our circular orbit hypothesis defined in Eq.(\ref{Eq:eqarlyBBH}) but also 
large enough so that no circumbinary structure exists yet.  
The design of our simulations, presented in the previous section, is consistent with an initial accretion disk whose extent exceeds the size of the Roche lobe of the primary black hole. \\
\noindent The temporal evolution of the gas density is displayed on Fig.\ref{fig:Density_evol} where one can see the outer part of the initial disk being swept away by the gravitational force exerted by the secondary black hole located outside of the simulation domain. The transition from the initial conditions to the shrunken accretion disk lasts for approximately one BBH period before stepping into a stabilization phase where the accretion disk slowly rearranges itself under the influence of the two black holes.

\noindent  It is noteworthy that the gravitational influence of the secondary black hole upon the circumprimary disk is also noticeable through the formation of overdense spiral arms. The outer parts of these arms rotate at the same angular velocity than the secondary black hole near the outer edge of the disk but  experience  the differential rotation of the gas as they progress through the inner part of the disk. The differential rotation of the gas leads to a spiral shape of the arms. The presence of these over-densities remains, however, throughout the whole simulation but in a more rolled up fashion making the spiral arms more prominent in the very inner part of the circumprimary disk.
\noindent Lastly,  we followed the disk evolution for more than $20$ periods of the binary and  we see on the last frame of Fig.\ref{fig:Density_evol} that the shrunken outer edge of the disk  is not circular anymore 
but shows eccentricity following the orbit of the secondary black hole.
\subsubsection{Assessing the outer radius of a circumprimary disk} 
\label{measurerout}
During the accretion disk evolution, precisely assessing the value of its  outer edge 
 is not straightforward as the shape of the outer edge evolves toward an elliptic structure. It is, however, useful to measure the azimuthally averaged 
 {outer radius in order to follow its evolution and final value under the presence of the secondary black-hole.}
   In order to do so we have adopted the following procedure to measure the radial extent of the disk: for any given snapshot, we first azimuthally average {the gas density of the disk so that we obtain its averaged 
   radial profile. This profile is displayed on the top of Fig.\ref{fig:Profile_vphi_rho} for our fiducial simulation where both initial and final profiles are represented. }\\
\noindent One can see that the density profile has evolved over time since the gas density has increased inside the disk while it has significantly dropped beyond a reference point where both initial and final densities match. We use this reference point to define the average outer edge of the disk $r_{\rm out}$ which corresponds to the radius where the gas density is equal to one tenth of the density at the reference point. This choice is motivated by the fact that a $90\%$ drop in the disk density results in a drop in thermal radiative emission power by a factor larger than $90\%$ depending on the gas equation of state. Such drop in thermal emission will be detectable compared to an extended disk and will be observationally interpreted as the outer accretion disk edge.  
It is noteworthy that the {azimuthally averaged} azimuthal velocity{, shown in the bottom of Fig.\ref{fig:Profile_vphi_rho},} remains very similar to the Keplerian velocity up to a radius close to
{the outer radius chosen with the method above. This}  is consistent with the gas no longer displaying the dynamical behavior of a balanced disk structure. Indeed, outside of $r_{\rm out}$, mass is very likely reacting to equivalent gravitational forces coming from the two objects composing the BBH.

\begin{figure}
    \centering
    \includegraphics[width=0.48\textwidth]{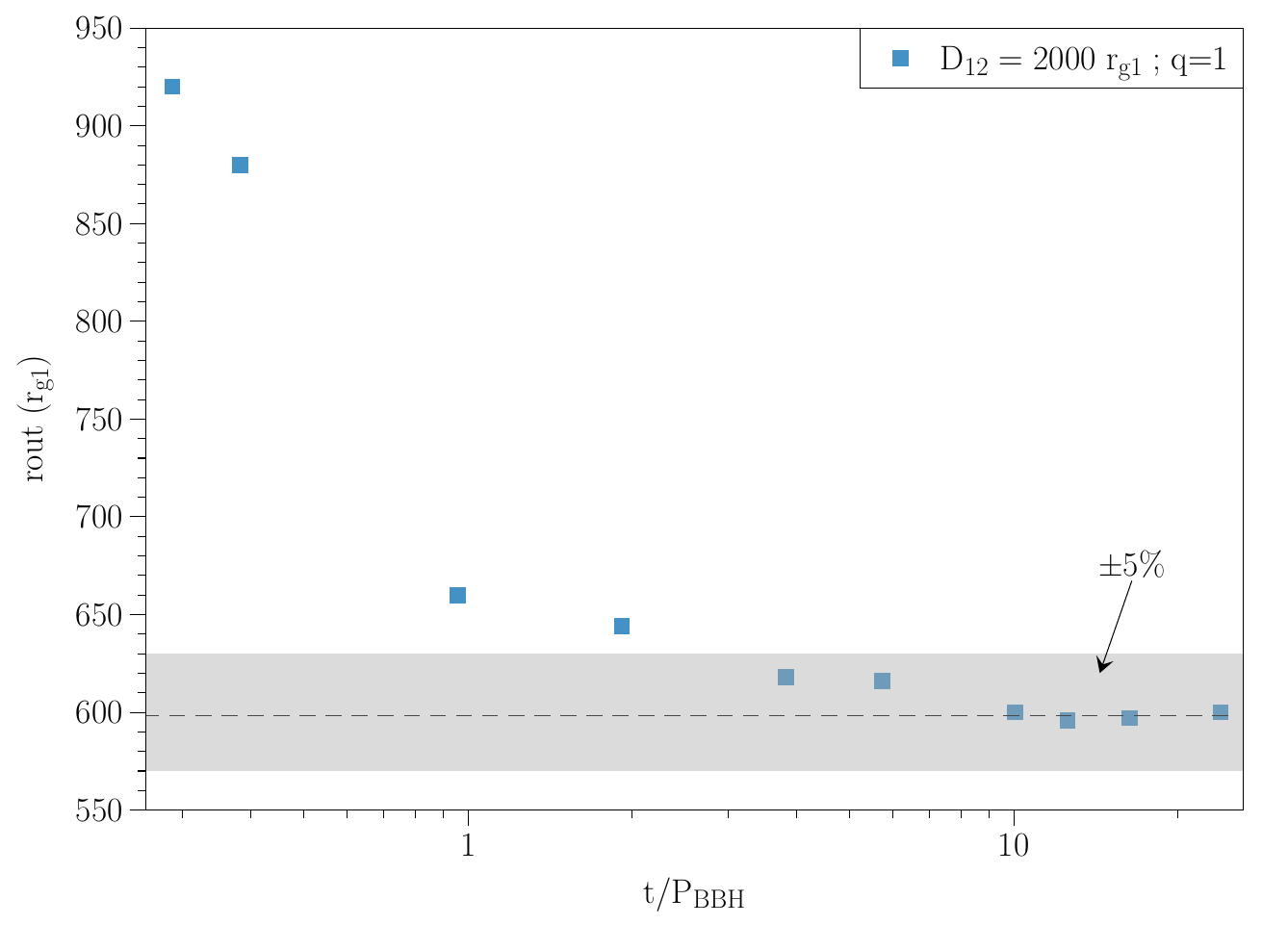}
    \caption{Temporal evolution of the averaged outer radius of the disk displayed in Fig.\ref{fig:Density_evol}. Time is normalized to the binary period $P_{\rm BBH}$. After a fast decrease from the initial outer edge value ($1.55\times 10^3\ r_{\rm g1}$), the outer edge radius stabilizes to a value of approximately $600\ r_{\rm g1}$. }
    \label{fig:rout_2000}
\end{figure}

\subsubsection{Evolution of the outer radius of the circumprimary disk} 

Applying this method to determine  $r_{\rm out}$ to our fiducial simulation leads us to Fig.\ref{fig:rout_2000} where the temporal evolution of $r_{\rm out}$ is displayed over nearly 25 $P_{\rm BBH}$. In such simulation, the initial disk whose outer edge lies at $r_{\rm out}=1500\ r_{\rm g1}$ (beyond the Roche lobe radius of the primary black hole) rapidly shrinks to a smaller extension which remains stationary after approximately $3P_{\rm BBH}$. The final outer edge radius of the disk in this simulation is then $\simeq 600\ r_{\rm g1}$. \\
The general behavior of $r_{\rm out}$ in this fiducial simulation matches the overall behavior of the disk in all our other simulations: a rapid decrease of the outer edge followed by a stabilization of the outer edge. In order to understand why the disk is experiencing such rapid shrinking in this simulation, one has to keep in mind that $P_{\rm BBH}$ is large compared to the gas dynamical time as $P_{\rm BBH}$ corresponds to $4$ orbiting periods of the gas near $r_{\rm out}$ and $6\times 10^3$ periods at the ISCO of the disk.
\subsubsection{Eccentricity of the outer edge of the circumprimary disk} 
\begin{figure}
    \centering
    \includegraphics[width=0.48\textwidth]{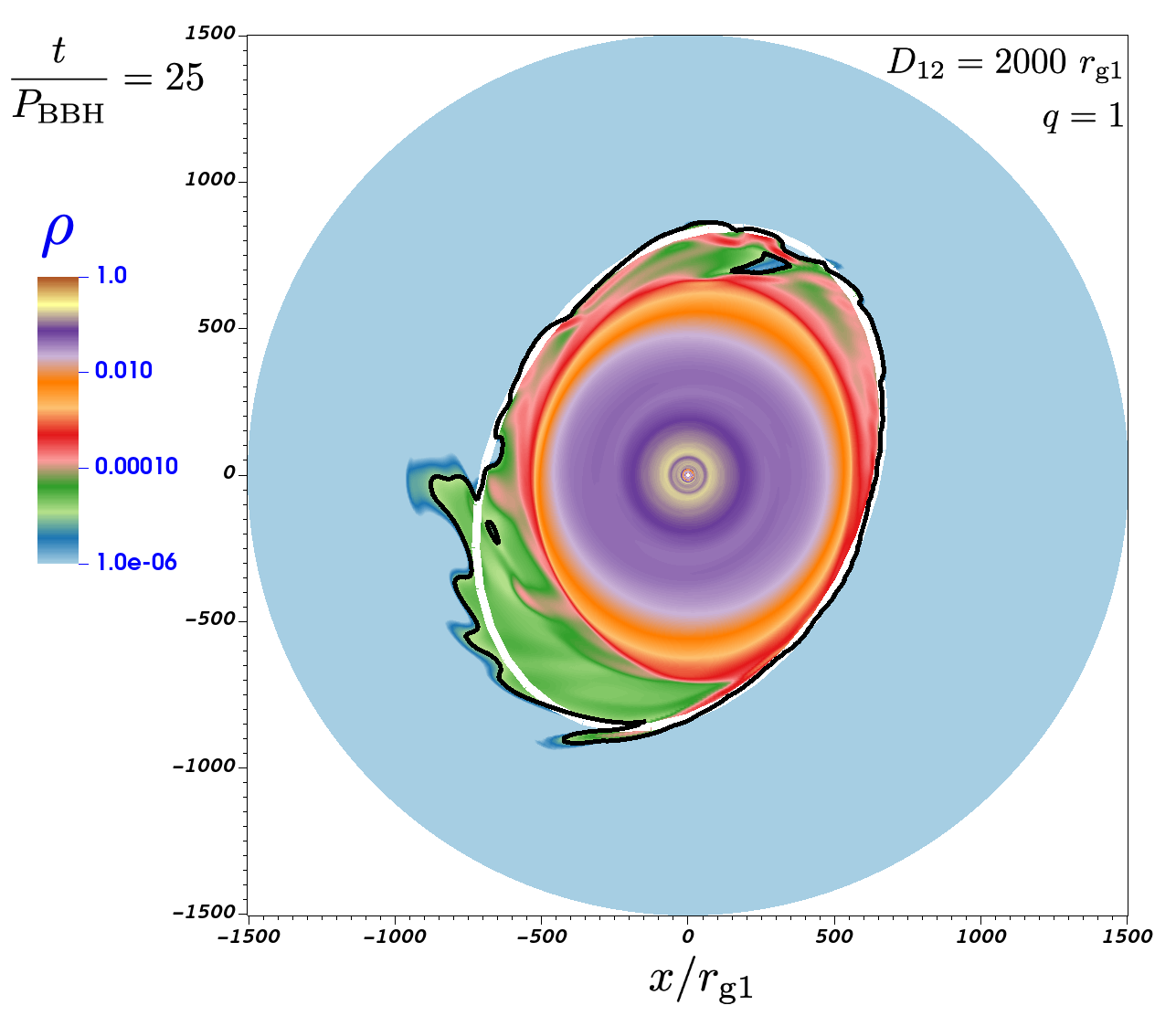} 
    \caption{Snapshot of the logarithmic density of the fiducial  simulation  (Fig.\ref{fig:Density_evol} $q=1$ and $D_{12}=2\times 10^3 r_{\rm g1}$) at the final stable stage. We have added a black solid line to indicate the location of the outer edge of the disk and a white solid line representing the elliptic fit of this outer edge whose eccentricity is $e=0.52 \pm 0.12$.}
    \label{fig:overlay_e}
\end{figure}
As we saw on Fig.\ref{fig:Density_evol} the final shape of the outer edge of the circumprimary disk
no longer exhibits an axisymmetric structure as the outer edge of the accretion disk appears to be elliptical.  Inferring the eccentricity of the outer part of the resulting disk was done using the following procedure: first, we calculate the contour of the gas density $\rho$ in the outer disk (where values of the density  are much smaller than unity but larger than the floor density).  
In a second step, we remove the spiral arms from this contour and finally we evaluate the approximation of this contour by an ellipse. The fitting algorithm is built from algebraic equations of an ellipse. All elliptical fits were obtained from the object-oriented tools for fitting conics and quadrics developed by \cite{MattJ23} within the Maple library.\\

\noindent For our fiducial case, displayed in Fig.\ref{fig:Density_evol}, we obtain a final eccentricity for the outer edge of the disk of $e=0.52 \pm 0.12$ (see Fig.\ref{fig:overlay_e}). The error bars are computed by doing the above procedure several times for 
different close parameters. This case has the largest error bar of all our simulations.

\subsection{Extension to all separations for equal-mass binary}

{In order to see how the outer edge of the disk behaves depending on the separation we have performed other equal-mass BBH simulations considering various separations
 $$D_{12}\in \{ 1, \ 1.5, \ 2, \ 3, \ 6\} \times 10^3\ r_{\rm g1}$$  which translate into physical separations between  
$[0.48,\  2.9] \times 10^{-4}\left(M_1/10^6M_\odot\right)$pc.}
In all those simulations we obtain a similar behavior to the one 
 displayed in Fig.\ref{fig:rout_2000} where we show the evolution of $r_{\rm out}$ as a function of time.

\begin{figure}
    \centering
    \includegraphics[width=0.48\textwidth]{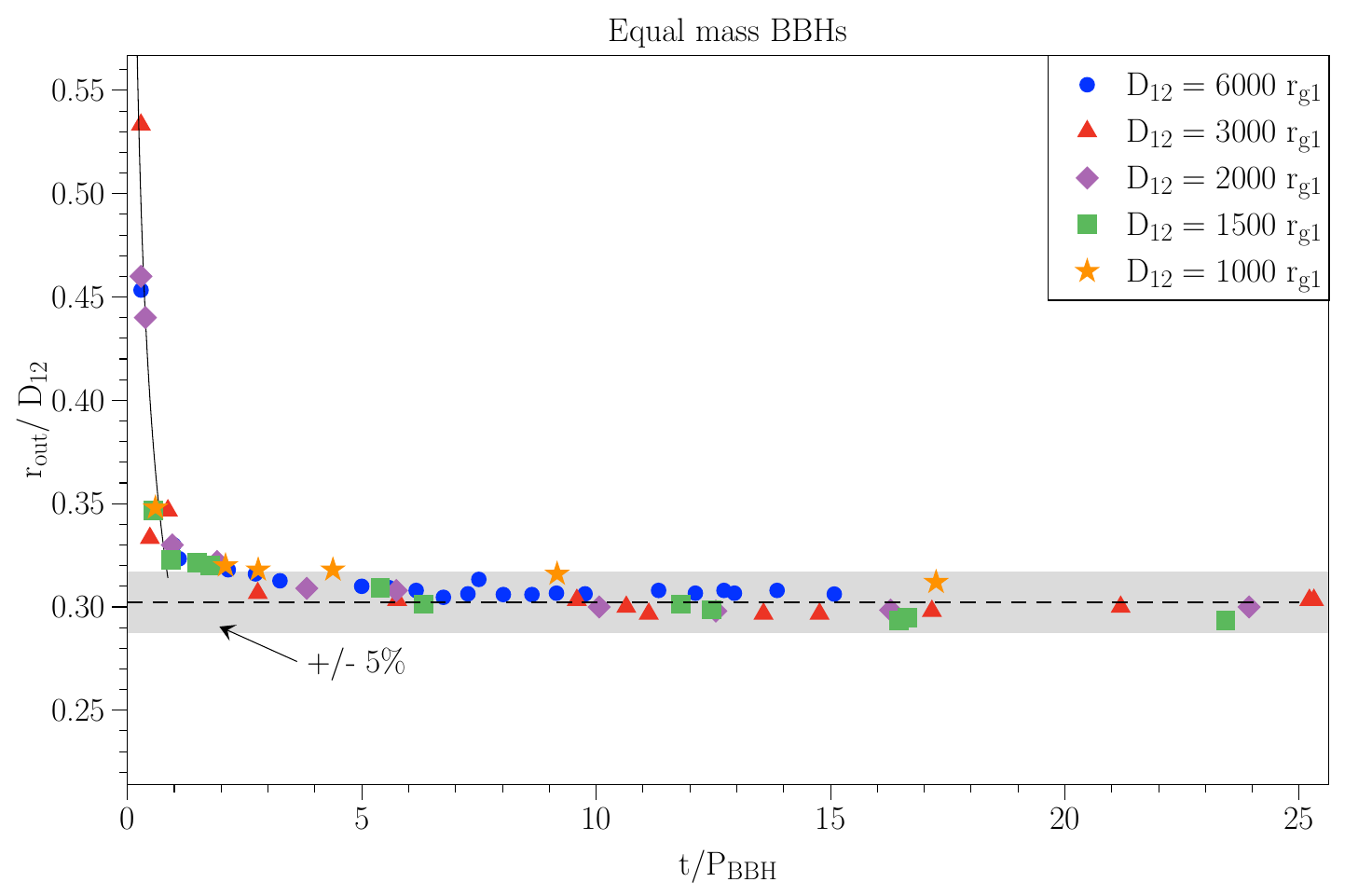}
    \caption{Temporal evolution of the outer edge of accretion disks in equal-mass binaries whose separation  range  from $500\ r_{\rm g1}$ to $6\times 10^3\ r_{\rm g1}$. The outer radius of the disk is normalized to the binary separation while time is normalized to the binary period $P_{\rm BBH}$.}
    \label{fig:rout_q=1}
\end{figure}

\noindent  In order to compare the outer edges of the circumprimary disk even though all the systems have different  separations, we have normalized each outer edge radius by {the respective separation of their binary}. 
As one can see on Fig.\ref{fig:rout_q=1} the {evolution} 
of $r_{\rm out}/D_{\rm 12}$ is not only very similar throughout our equal-mass BBH sample with a rapid drop of the outer radius occurring
 over a few $P_{\rm BBH}$,  but they also all converged toward a final value of $r_{\rm out}\simeq (0.3 \pm 0.015) D_{12}$. 
This is not unexpected as the dynamical aspect of the equal mass binary system scales with the distance.
We have also computed the eccentricity associated with those different equal-mass BBHs and found that they are all coherent with $e= 0.62^{+0.13}_{-0.22}$. \\
\noindent The results from the equal-mass BBH simulations  suggest that the binary separation has {little to} no impact on the values of $r_{\rm out}/D_{12}$ and the eccentricity of the outer edge. In order to investigate if the BBH mass ratio $q$ has a significant impact on those quantities we need to allow the mass ratio to change and see how that impacts the overall behavior of the outer edge and its eccentricity.

\begin{table*}[htbp]
\caption{BBH mass ratio $q$ and black hole separation $D_{12}$ for our set of hydrodynamical simulations.}
\centering                        
\begin{tabular}{c|cccccc}     
\hline\hline               
$\displaystyle\frac{D_{12}}{r_{\rm g1}}$ & $500$ & $10^3$ & $1.5\times 10^3$ & $2\times 10^3$ & $3\times 10^3$ & $6\times 10^3$\\         
\hline                      
   $q$ & $9\times 10^{-4}$ & $1 ; 0.3 ; 0.05 ; 0.005$ & $1 ; 0.3 ; 0.05 $ & $1 ; 0.3 ; 0.05$ & $1 ; 0.3 ; 0.1 ; 0.01$ & $1$\\
\hline\hline                                  
\end{tabular}
\tablefoot{The separation is expressed in units of the primary black hole gravitational radius $r_{\rm g1}$. Our simulation sample exhibits binary separation ranging from $499.55\ r_{\rm g}$ to $3\times 10^3r_{\rm g}$  where $r_{\rm g}$ is the binary gravitational radius defined as $r_{\rm g}=(1+q)r_{\rm g1}$.}               
\label{Tab_all}    
\end{table*}

\section{Accretion disk outer edge in pre-CBD BBH: non equal-mass binaries versus separation}

 In order to go beyond the equal-mass BBH framework we have performed a series of simulations where we have considered six different BBH mass ratios $q$ and six separations $D_{12}$ to get a 
 good representation of the parameter space and see how the outer radius behaves for all cases. 
 The range of mass ratios considered in our simulation sample is $q\in [9\times 10^{-4},1]$. The lower value of this interval was constrained by computational time limitations while we set the upper limit to $q=1$ 
 as we run the simulation around the most massive black-hole. \\
 The range of binary separation in our sample is $D_{12}/r_{\rm g1} \in [500,6\times 10^3]$. The upper limit of the black hole separation corresponds to an accretion disk whose initial outer edge is consistent with a typical size of accretion disks observed around supermassive black holes (see e.g. \citealt{Jha21} and references therein). The lower limit corresponds to a set-up where it is numerically possible to address simulations with very low mass ratio $q$ down to~$\sim 10^{-3}$. Indeed at such low value of $q$, one has to perform long-term simulations over nearly one thousand 
 BBH periods in order for the disk to reach an equilibrium state. Such long-term simulations are then easier to perform if the separation is small but still larger than the limit for the circular orbit condition (Eq.\ref{Eq:eqarlyBBH}) and coherent with a pre-CBD BBH. 
\subsection{Measurement of the outer radius of circumprimary disks in non-equal mass pre-CBD BBH}
\begin{figure}
    \centering
     \includegraphics[width=0.48\textwidth]{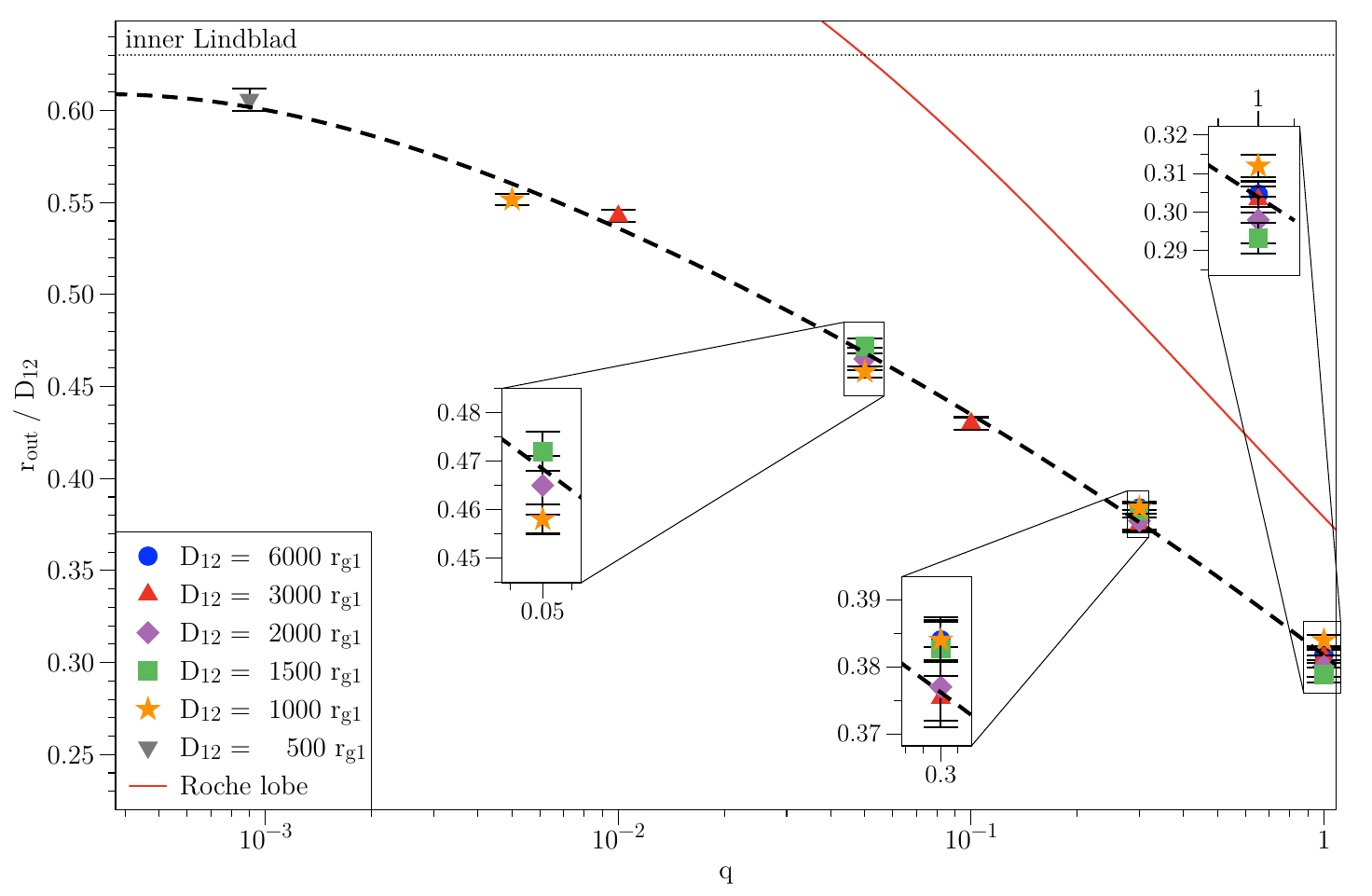}
    \caption{Ratio of the average disk outer radius $r_{\rm out}$ to the BBH separation $D_{12}$ as a function of the BBH mass ratio $q$. The colored points correspond to 
    various BBH separations. For any given BBH mass ratio, the values of $r_{\rm out}/D_{12}$ are very close for all separations showing that this ratio does only depend on the 
    BBH mass ratio $q$. One can also notice that all values of $r_{\rm out}$ are smaller than both the size of the Roche lobe \citep{Eggleton83} and the location of the innermost Lindblad resonance as expected for any stable accretion disk.}
    \label{fig:rout_vs_q_final}
\end{figure}
For each simulation {listed in Tab.\ref{Tab_all}} we measured the outer edge of the accretion disk in the exact same way than in Sect. \ref{measurerout} and compiled the results in order to study the relationship between $r_{\rm out}$ and the two BBH parameters $q$ and $D_{12}$.
Fig.\ref{fig:rout_vs_q_final} represents the value of the ratio $r_{\rm out}/D_{12}$ as a function of the mass ratio parameter $q$. The main result provided by this figure is that the ratios $r_{\rm out}/D_{12}$ are remarkably similar for any given mass ratio $q$. 
Indeed, for any given mass ratio, the final outer edge radius in each simulation is equal to the same fraction of $D_{12}$, whatever is the BBH separation $D_{12}$. Scanning the various values of BBH mass ratio, we find that this fraction is however dependent on $q$ as its value increases as $q$ decreases. 
In order to check the consistency of our simulations, we have represented on Fig.\ref{fig:rout_vs_q_final} the distance from the primary black hole to the $L_1$ Lagrange point of the BBH system as well as the location of the inner Lindblad resonance.  All the values of $r_{\rm out}$ found in our sample are smaller than these two distances proving the stability of the final structure of the accretion flow around the primary black hole of the BBH. At high mass ratios, the location of the $L_1$ is the main limitation while at low mass ratios, the Lindblad inner resonance is dominating the extent of the circumprimary disk. We provide here an empirical formula{, represented by the dash line on Fig.\ref{fig:rout_vs_q_final},} that fits the results arising from our  simulations as 
\be
\frac{r_{\rm out}}{D_{12}} = -0.18q^{0.044}\log_{10}{\left(0.019q+6.8\times 10^{-6}\right)}
\label{Eq:Fit_rout}
\ee
The coefficients of this formula were derived by minimizing a standard $\chi^2$ test based on the values provided by our simulations (with a $\chi^2=0.99$ and residues below $1\%$). From Eq.\ref{Eq:Fit_rout}. we see that the outer radius of the circumprimary disk stays proportional to binary separation, with the proportionality factor only depending on the mass ratio. 
This means that the outer radius of the circumprimary disk is a reliable indicator of the binary separation during the pre-CBD phase of the inspiral. When comparing our results with \citet{Pichardo05}, we find a general agreement in terms of mass ratio dependance while obtaining marginally larger outer radii. We interpret that slight difference to originate from different definitions of the outer radius: it is defined  from the outermost non-intersecting test-particle loop at the
binary periastron in \citet{Pichardo05} while we compute an average value of the outer radius based on the drop of the thermal emissivity from the disk. \\ 
The result presented here, linking any evolution of the outer edge of the disk with a decrease in separation, is interesting not only as a test for all the existing BBH candidates, but also
 to look for new potential candidates. Indeed, this relation emphasizes the need to 
 question any AGN system exhibiting evidence of an accretion disk with a smaller outer radius than expected.  
 In our sample we covered three orders of magnitude for the $q$ parameter which leads to $r_{\rm out} \in \{0.3, \ 0.6\} D_{12}$. 
 Taking those numbers into account provides a location range for the search of the potential secondary black hole  
{at the origin of this shaved outer disk}.
\begin{figure}
    \centering
    \includegraphics[width=0.48\textwidth]{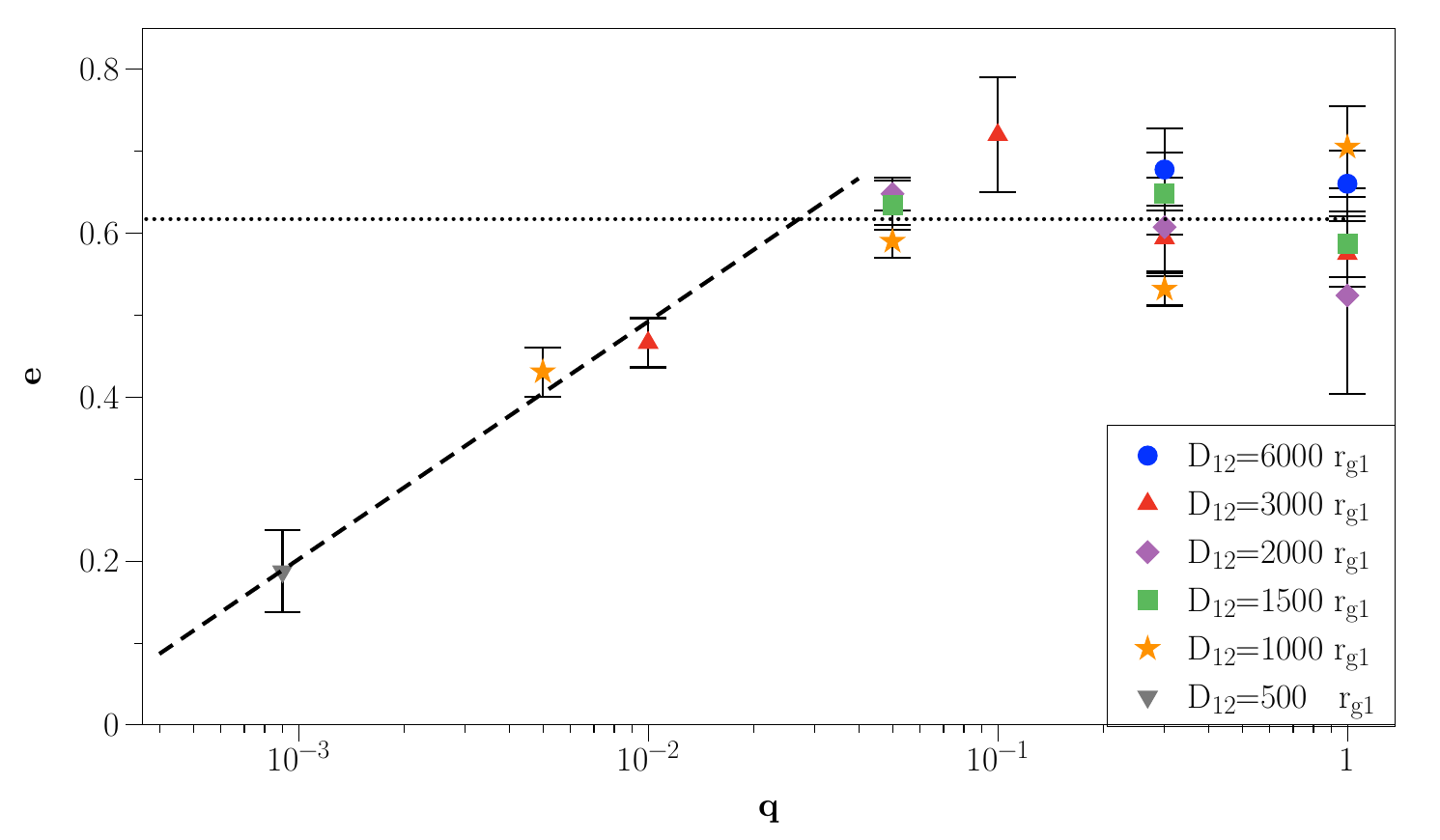}
    \caption{Eccentricity of the outer disk as a function of the binary mass ratio $q$. The eccentricity mostly depends on the binary mass ratio parameter $q$ and very little on the binary separation $D_{12}$. }
    \label{fig:e_vs_q}
\end{figure}
\subsection{Eccentricity of the outer edge of a  circumprimary disk in non equal-mass pre-CBD BBH}
\label{sec:e}

Following the same technique presented in Sect. \ref{equalBBH} we looked at the eccentricity of the outer edge of the circumprimary disk in all our simulations.
The results obtained from the fitting of the shape of the outer edge of the circumprimary disk  are displayed on Fig.\ref{fig:e_vs_q} along with the associated uncertainties. As expected we can see that for the smallest values of the mass ratio $q$, the eccentricity $e$ is small and increases with $q$. 
On the other hand we find that for mass ratio $q>10^{-2}$  the outer edge of the disk exhibits a high eccentricity of $e\sim 0.6$ as can be seen in the final snapshots of Fig.\ref{fig:Density_evol} and in Fig.\ref{fig:overlay_e}. 
As a result, the outer disk eccentricity cannot be used as a potential tracer of the mass ratio for $q<10^{-2}$.
On top of that, we also need to keep in mind that we focused on black-hole orbits without eccentricity themselves.  Using a purely newtonian 2D hydrodynamical approach \cite{HD100546} 
showed that increasing the eccentricity of the gravitational disruptor tends to form a more circular looking outer edge of the disk though no precise study of the outer edge eccentricity was performed.
If such effects also apply to BBH, it would mean that a non-detection of eccentricity in the outer disk could be the result of an eccentric binary and not only a small mass ratio.
Nevertheless, if through other means, such as gravitational wave detection, we have access to the binary eccentricity, then the shape of the outer disk could further discriminate between low and high mass ratios.

\section{Conclusions}

 In this paper we considered binary black hole systems in the pre-circumbinary stage where the separation between the two black holes can be larger than the typical size of AGN disks. We thus assumed that during this pre-CBD stage, the most massive black hole may still be surrounded by a pre-existing accretion disk dubbed as the circumprimary disk. We here studied how the presence of a gravitationally bound secondary black hole impacts the evolution of the circumprimary  disk in a BBH system with the aim to explore if any characteristic features 
of this disk could  be linked with the binary parameters, thus having the potential to help identification of BBH.
In this first step we chose to focus  on systems whose orbital separation is large (up to a few thousands gravitational radii) so that it still belongs to the pre-CBD disk BBH stage.  
This choice was motivated  by the fact that it is a relatively unstudied stage of pre-merger systems that rules over the formation of the circumbinary disk prior to the merger phase. Describing such early stage of the BBH is one of the keys to accurately picture the evolution of BBH systems. In pre-CBD systems, the black-hole inspiral motion is very slow so it enables us to consider the two black holes to move on circular orbits while keeping their separation constant.\\

\noindent  In order to link the circumprimary disk characteristics with the binary parameters we performed a series of 2D hydrodynamical simulations for six separations covering the pre-CBD BBH stage
and six mass ratios ranging over three orders of magnitude. As we focused on the outer edge of the circumprimary disk, we performed those simulations in a classical, pseudo-newtonian, framework 
which allowed us to follow the system from more than twenty orbits of the binary for the largest separation to almost a thousand for the closest separation.
In each simulation we reached a stable state with the overall  size and shape of the outer circumprimary disk remaining constant (see Fig.\ref{fig:rout_q=1}). This allows us
to use  the end result of those simulations as an \lq instantaneous snapshot\rq\  of a pre-CBD system with a given separation, hence a given time to merger.\\

\noindent  
 We found that the presence of a secondary black hole has three main effects upon the circumprimary accretion disk, similarly to previous studies performed on other type of binary systems:
\begin{description}
\item[\tt - Eccentric outer disk.] As visible on  final snapshots of Fig.\ref{fig:Density_evol}, the outer circumprimary disk becomes elliptical with  $e= 0.62^{+0.13}_{-0.22}$ when both $q>0.05$ and the binary is in a 
circular orbit. Because of those two requirements, the eccentricity of the primary outer disk is not a good tracer for the mass ratio of the binary. Nevertheless, it can be used as an extra check for binary black hole 
candidates whose parameters, including eccentricity, have been deduced from another method. 

\item[\tt - Spiral arms.] Another effect  is the  existence of two overdense spiral arms sweeping through the circumprimary disk following the orbit of the secondary black hole. 
Spiral arms being quite common features of accretion disks, they cannot be used to identify BBH. On top of that, they tend to be of low amplitude once the system is in a steady state, making them 
not a reliable test for existing BBH candidates.

\item[\tt - Truncated outer disk.] Finally, the most valuable consequences of the presence of a secondary black hole is the significant truncation of the outer circumprimary disk. 
  Going one step further, we were able to express the position of that truncated outer edge as function of the binary parameter: $r_{\rm out} = f(q) D_{12}$\footnote{
  $r_{\rm out} = -0.18\ q^{0.044} \log_{10}{\left(0.019q+6.8\times 10^{-6}\right)} \ D_{12} $}, hence linking directly the truncation of the disk to the time-to-merger. 

 \end{description}
 \noindent  The presence of a smaller than expected disk around a super massive black hole (SMBH) is not specific to the presence of a gravitational disruptor in the system. 
 Indeed, it could be related to a gas-poor environment preventing the black-hole from having an extended disk which could be inferred from other observations of the region,  
 or it could have been truncated by a fly-by, an unbound massive object,  in that case the resulting outer radius would not only be different, but the secondary object could be detectable \citep{HD100546}. 
 Hence, while not unique to pre-CBD BBHs, the existence of a truncated accretion disk around a SMBH can be considered a flag  implying the need to further study the source to test
 for a potential binary companion.\\
 \noindent This is of particular interest as \cite{McHardy23} recently published the \lq first detection of the outer edge of an AGN accretion disk\rq\  in NGC 4395. If a gravitationally bound secondary black hole is responsible for such small accretion disk, our results predict a separation of two black holes to range from $1.7$ to $3.3$ times the outer radius of the accretion disk provided $10^{-3}\leq q\leq 1$. 
 Assessing the detectability of these features is not in the scope of this paper and we refer the reader to a companion paper devoted to the observational implications of the findings of the present paper.

\begin{acknowledgements}

The numerical simulations we have presented in this paper were produced on the DANTE platform (AstroParticule \& Cosmologie, France).
Part of this study was supported by the LabEx UnivEarthS, ANR-10-LABX-0023 and ANR-18-IDEX-0001. The data that support the findings of this study are available from the corresponding author, FC, upon request.
\end{acknowledgements}

\bibliographystyle{aa}
\bibliography{Scuplting_Outer_Edge}

\end{document}